\documentclass[submission,copyright,creativecommons,unicode,UKenglish]{eptcs}
\providecommand{\event}{CL\&C'16} % Name of the event you are submitting to
\usepackage[utf8]{inputenc}
\usepackage[UKenglish]{babel}
\usepackage{xparse}

\usepackage{amsmath,amssymb,latexsym}
\usepackage{amsthm}
\usepackage{thmtools}
\usepackage{bm}

\usepackage[pdftex]{virginialake}
\usepackage{graphics}

\DeclareDocumentCommand{\Seq}{m}{\langle{#1}\rangle}
\newcommand{\emptyseq}{\Seq{}}

\newcommand*{\size}[1]{\lvert{#1}\rvert}

\declaretheorem[numberwithin=section]{theorem}
\declaretheorem[sibling=theorem]{lemma}

\newcommand{\Gram}[1]{\mathcal{#1}}

\renewcommand*{\vec}[1]{\bm{#1}}
\newcommand{\con}{\textsf{c}}
\newcommand{\perm}{\textsf{p}}
\newcommand{\weak}{\textsf{w}}
\newcommand{\cut}{\textsf{cut}} 

\vlnostructuressyntax
\newdimen\myvltreesize
\renewcommand{\vltrauxx}[5]{\vldaux
   {#1}
   {#2}
   {\vlhy{#3}}
   {}
   {\hbox{$\vcenter{\xy
          0;<\hsize,0pt>:<0pt,#4\hsize>::
          (-0.5,0.5);(0.5,0.5)**\crv{(0.1,0.6)&(-0.1,0.4)};
          (0,0)**@{-};(-0.5,0.5)**@{-};
          (0,0.293)*{#5\strut}
          \endxy}$}}
   {\kern\deropen}}
\newcommand{\vlstrf}[4]{\vltrf{\hbox{\small$#1$}}{#2}{\vlshy{}}{#3}{\vlshy{}}{#4}}
\newcommand{\vlbtrf}[4]{\vlstrf{#1}{#2}{\vlshy{\hskip#3}}{#4}}
\newcommand{\vlhtr}[2]{\vlbtrf{#1}{#2}{2em}{1}}
\newcommand{\vlshy}[1]{\vlhyaux{$#1$}}

\title{On the Herbrand content of LK}
\author{Bahareh Afshari
	\qquad\qquad
	Stefan Hetzl
	\qquad\qquad
	Graham E.\ Leigh
	\institute{TU Wien,
		Austria}
	\email{\{bahareh.afshari,stefan.hetzl,graham.leigh\}@tuwien.ac.at}
}
\def\titlerunning{On the Herbrand content of LK}
\def\authorrunning{B. Afshari, S. Hetzl \& G.E. Leigh}

\begin{document}
%%=========================
\maketitle
\begin{abstract}
	We present a structural representation of the Herbrand content of LK-proofs with cuts of complexity prenex $\Pi_2/\Sigma_2$. The representation takes the form of a typed non-deterministic tree grammar $\mathcal G$ of order $2$ which generates a finite language, $L(\mathcal G)$, of first-order terms that appear in the Herbrand expansions  obtained through  cut-elimination.
	In particular, for every Gentzen-style reduction $\pi\rightsquigarrow \pi'$ between LK-proofs we study the induced grammars,  respectively $\mathcal G$ and $\mathcal G'$, and classify the cases in which language equality, $L(\mathcal G)=L(\mathcal G')$, and  language inclusion, $L(\mathcal G)\supseteq L(\mathcal G')$, hold.
\end{abstract}
%%=========================
\section{Introduction}\label{sec:introduction}
%%=========================
In classical first-order logic a proof can be considered as being composed of two layers: on the one hand the terms by which quantifiers are instantiated, and on the other hand, the propositional structure. 
This separation is most clearly illustrated by Herbrand's theorem~\cite{Her30,Bus95}: a formula is valid if and only if there is a finite expansion (of existential quantifiers to disjunctions and universal quantifiers to conjunctions of instances) which is a propositional tautology. Such Herbrand expansions can be transformed to and obtained from cut-free sequent calculus proofs in a quite straightforward way.

It is non-trivial to formally extend this separation to proofs with cuts.
An approach which has been successful in this respect is the use of tree grammars, introduced in~\cite{Het12} for proofs with $\Pi_1$-cuts and extended to $\Pi_2$-cuts in~\cite{AHL15,AHL16}.
In this setting, a proof in sequent calculus induces a tree grammar which bears all instances of the end-sequent as well as the instantiation structure of the cuts without direct reference to the cut formulæ themselves:
one obtains a Herbrand expansion by computing the language of the grammar.

In addition to the proof-theoretic interest behind an abstract representation of proofs with cut, proof grammars provide a number of applications.
Motivated by the aim to structure and compress automatically generated proofs, an algorithm for cut-introduction based on proof grammars has been developed in~\cite{HLW12,HLRW14}.
This method has been implemented and empirically evaluated with good results in~\cite{HLRTW14}.
An extension of these techniques to the case of
proofs with $\Pi_1$-induction has led to a new technique for inductive theorem proving~\cite{EH15Inductive} which is currently being implemented.
A final application of proof grammars is in the area of proof complexity, where lower bounds on the length of proofs with cuts (which are notoriously difficult to control) are obtained by transferring lower bounds on the size of the corresponding grammar~\cite{Eberhard15Compressibility,EberhardXXCompressibility}.

There are other formalisms which allow Herbrand expansions to be computed in a way that abstracts from the propositional
structure.
The historically first such formalism is Hilbert's $\varepsilon$-calculus~\cite{HB39}.
In~\cite{GK05} Gerhardy and Kohlenbach adapt Shoenfield's variant of G\"odel's Dialectica interpretation to a system of pure predicate logic.
Recent work, related to proof nets, is that of Heijltjes~\cite{Hei10} and McKinley~\cite{McK13}, and a similar approach, in the formalism of expansion trees~\cite{Mil87}, can be found in~\cite{HW13}.
What sets proof grammars apart from these formalisms is that they not only compute Herbrand expansions but provide a (well-understood) abstract description of its structure which is crucial for the applications mentioned above.

In the present paper we provide an intermediate formalism between proof grammars and functional interpretations with the aim of studying the relationship between the two approaches.
This intermediate formalism is presented as a grammar but instead of capturing the instantiation structure directly, it is given by a brief line-by-line definition on the proof, as functional interpretations usually are.
The necessity for computing more than one witness (reflected by the case distinction constants in the Gerhardy--Kohlenbach version of the Dialectica interpretation~\cite{GK05}) is reflected by non-deterministic production rules in the grammar.

 The main result we prove in this paper is stated below. Note that in the presence of  Skolemisation it suffices to  consider proofs with $\Sigma_1$ end sequents.
\begin{theorem}\label{thm:Main_Thm}
Let $\pi$ be a proof of  $\exists \vec v F$ with $F$ quantifier-free in which cut-formul\ae{} are prenex $\Pi_2$ or $\Sigma_2$.
There exists an acyclic context-free grammar $\mathcal G$ such that
$\bigvee_{\vec t\in L(\Gram G)} F(\vec t)$ is valid.
Moreover, $L(\Gram G)$ contains the Herbrand set extracted from any cut-free proof that can be obtained from $\pi$ via a sequence of cut reductions (see~\autoref{fig:reductionsteps}) that always reduces to the weak (quantifier) side of a cut before the strong side.

More generally, $L(\mathcal G)$ covers the Herbrand set of any cut-free proof obtained from $\pi$ by a sequence of reductions fulfilling the following two restrictions.
\begin{enumerate}
  \item A contraction on a universally quantified $\Pi_2$ formula is reduced only when no other reduction rule is applicable (to this cut);
  \item If two cuts are permuted in the form
  \begin{gather*}
	\vlderivation
	{
	  \vliin{\cut}{}{\Gamma,\Delta,\Lambda}
	  {
		\vliin{\cut}{}{B, \Gamma,\Delta}
		{
		  \vlhy{A,B,\Gamma}
		}{
		  \vlhy{\bar A,\Delta}
		}
	  }{
		\vlhy{\bar{B},\Lambda}
	  }
	}
	\qquad\rightsquigarrow\qquad
	\vlderivation
	{
	  \vliin{\cut}{}{\Gamma,\Delta,\Lambda}
	  {
		\vliin{\cut}{}{A,\Gamma,\Lambda}
		{
		  \vlhy{A,B,\Gamma}
		}{
		  \vlhy{\bar{B},\Lambda}
		}
	  }
	  {
		\vlhy {\bar A,\Delta}
	  }
	}
  \end{gather*}
  then one of $A$ and $B$ is not $\Pi_2$.
\end{enumerate}
\end{theorem}
%%=========================
\section{The system LK}\label{sec:LK_and_cut_reductions}
%%=========================
\begin{figure}[!b]
\centering
\fbox{
	\begin{minipage}{11 cm}
	Axioms:\quad\quad\quad\quad \quad
$\vlderivation{
        \vlhy{A,\bar A}}$	\quad\quad for $A$ an atomic formula\smallskip

	\noindent
	Inference rules:
	\begin{tabular}{@{\quad}c@{\quad\quad}c@{\quad\quad}c}
 $\vlderivation{
   \vlin{\lor}{}{A\lor B, \Gamma}{
          \vlhy{A,B, \Gamma}}}$
	& 
	
$\vlderivation{
  \vliin{\land}{}{A\land B,\Gamma,\Delta}{
          \vlhy{A,\Gamma}}{
          \vlhy{B,\Delta}}}$
	\\[1.5em]
	$\vlderivation{
	  \vlin{\forall}{}{\forall v A,\Gamma}{
          \vlhy{A(v/\alpha),\Gamma}}}$	
	&
	$\vlderivation{
	  \vlin{\exists}{}{\exists v A,\Gamma}{
          \vlhy{A(v/t),\Gamma}}}$	
	&
	$\vlderivation{
	  \vliin{\cut}{}{\Gamma,\Delta}{
          \vlhy{A,\Gamma}}{
          \vlhy{\bar A,\Delta}}}$
	\\[1.5em]
	$\vlderivation{
	  \vlin{\weak}{}{A,\Gamma}{
          \vlhy{\Gamma}}}$
	&
	$\vlderivation{
	  \vlin{\con}{}{A,\Gamma}{
          \vlhy{A,A,\Gamma}}}$
          &
	$\vlderivation{
	  \vlin{{\perm}}{}{\Gamma,A, B,\Delta}{
          \vlhy{\Gamma,B,A,\Delta}}}$
	%\\[1.5em]
	\end{tabular}
	\end{minipage}
}
\caption{Axioms and rules of LK. The usual eigenvariable conditions for quantifier introduction apply.}
\label{tab:calculus}
\end{figure}
\begin{figure}[p!]
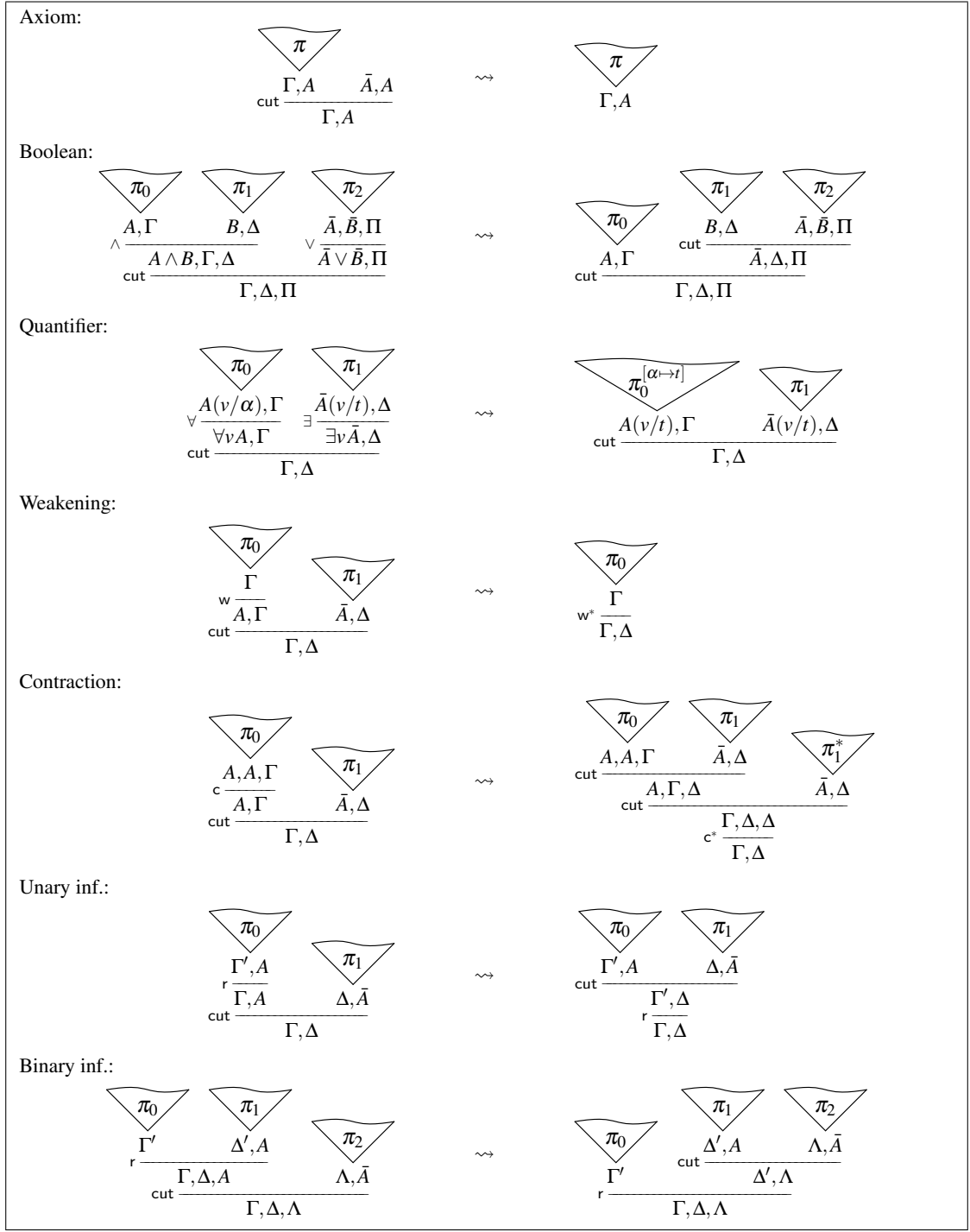
%
\centering
\fbox{
	\begin{minipage}{0.9\textwidth}
\footnotesize\abovedisplayskip=0pt\belowdisplayskip=2pt%
Axiom:
\begin{align*}
  \vlderivation
  {
	\vliin{\cut}{}{\Gamma,A}
	{
	  \vlhtr{\pi}{\Gamma,A}
	}{
	  \vlhy{\bar{A},A}
	}
  }
&&\rightsquigarrow&&&
  \vlderivation
  {
	\vlhtr{\pi}{\Gamma,A}
  }
\intertext{Boolean:}  %  BOOLEAN
  \vlderivation
  {
	\vliin{\cut}{}{\Gamma,\Delta,\Pi}
	{
	  \vliin{\land}{}{A\land B,\Gamma, \Delta}
	  {
		\vlhtr{\pi_0}{A,\Gamma}
	  }{
		\vlhtr{\pi_1}{B,\Delta}
	  }
	}{
	  \vlin{\lor}{}{\bar{A}\lor \bar{B},\Pi}
	  {
		\vlhtr{\pi_2}{\bar{A},\bar{B},\Pi}
	  }
	}
  }
&&\rightsquigarrow&&&
  \vlderivation
  {
	\vliin{\cut}{}{\Gamma,\Delta,\Pi}
	{
	  \vlhtr{\pi_0}{A,\Gamma}
	}{
	  \vliin{\cut}{}{\bar{A},\Delta, \Pi}
	  {
		\vlhtr{\pi_1}{B,\Delta}
	  }{
		\vlhtr{\pi_2}{\bar{A},\bar{B},\Pi}
	  }
	}
  }
\intertext{Quantifier:} % QUANTIFIER
  \vlderivation
  {
	\vliin{\cut}{}{\Gamma,\Delta}
	{
	  \vlin{\forall}{}{\forall v\, A,\Gamma}
	  {
		\vlhtr{\pi_0}{A(v/\alpha),\Gamma}
	  }
	}{
	  \vlin{\exists}{}{\exists v\, \bar{A},\Delta}
	  {
		\vlhtr{\pi_1}{ \bar{A}(v/t),\Delta}
	  }
	}
  }
&&\rightsquigarrow&&&
  \vlderivation
  {
	\vliin{\cut}{}{\Gamma,\Delta}
	{
	  \vlbtrf{\pi_0^ { [\alpha\mapsto t] } }{A(v/t),\Gamma}{6em}{0.6}
	}{
	  \vlbtrf{\pi_1}{\bar{A}(v/t),\Delta}{2em}{1}
	}
  }
\intertext{Weakening:} % WEAKENING
  \vlderivation
  {
	\vliin{\cut}{}{\Gamma,\Delta}
	{
	  \vlin{\weak}{}{A,\Gamma}
	  {
		\vlhtr{\pi_0}{\Gamma}
	  }
	}{
	  \vlhtr{\pi_1}{\bar{A},\Delta}
	}
  }
&&\rightsquigarrow&&&
  \vlderivation
  {
	\vlin{\weak^*}{}{\Gamma,\Delta}
	{
	  \vlhtr{\pi_0}{\Gamma}
	}
  }
\intertext{Contraction:} %CONTRACTION
  \vlderivation
  {
	\vliin{\cut}{}{\Gamma,\Delta}
	{
	  \vlin{\con}{}{A,\Gamma}
	  {
		\vlhtr{\pi_0}{A,A,\Gamma}
	  }
	}{
	  \vlhtr{\pi_1}{\bar A,\Delta}
	}
  }
&&\rightsquigarrow&&&
  \vlderivation
  {
	\vlin{\con^*}{}{\Gamma,\Delta}
	{
	  \vliin{\cut}{}{\Gamma,\Delta,\Delta}
	  {
		\vliin{\cut}{}{A,\Gamma,\Delta}
		{
		  \vlhtr{\pi_0}{A,A,\Gamma}
		}{
		  \vlhtr{\pi_1}{\bar A, \Delta}
		}
	  }{
		  \vlhtr{\pi_1^*}{\bar A,\Delta}
	  }
	}
  }
\intertext{Unary inf.:} % UNARY INFERENCE
  \vlderivation
  {
	\vliin{\cut}{}{\Gamma,\Delta}
	{
	  \vlin{\textsf r}{}{\Gamma,A}
	  {
		\vlhtr{\pi_0}{\Gamma',A}
	  }
	}{
	  \vlhtr{\pi_1}{\Delta,\bar{A}}
	}
  }
&&\rightsquigarrow&&&
  \vlderivation
  {
	\vlin{\textsf r}{}{\Gamma,\Delta}
	{
	  \vliin{\cut}{}{\Gamma',\Delta}
	  {
		\vlhtr{\pi_0}{\Gamma',A}
	  }{
		\vlhtr{\pi_1}{\Delta,\bar{A}}
	  }
	}
  }
\intertext{Binary inf.:} % BINARY INF
  \vlderivation
  {
	\vliin{\cut}{}{\Gamma,\Delta,\Lambda}
	{
	  \vliin{\textsf r}{}{\Gamma,\Delta,A}
	  {
		\vlhtr{\pi_0}{\Gamma'}
	  }{
		\vlhtr{\pi_1}{\Delta',A}
	  }
	}{
	  \vlhtr{\pi_2}{\Lambda,\bar{A}}
	}
  }
&&\rightsquigarrow&&&
  \vlderivation
  {
	\vliin{\textsf r}{}{\Gamma,\Delta,\Lambda}
	{
	  \vlhtr{\pi_0}{\Gamma'}
	}{
	  \vliin{\cut}{}{\Delta',\Lambda}
	  {
		\vlhtr{\pi_1}{\Delta',A}
	  }{
		\vlhtr{\pi_2}{\Lambda,\bar{A}}
	  }
	}
  }
\end{align*}
\end{minipage}
}
\caption{One-step cut reduction and permutation rules.
	In the final two reductions, \textsf{r} denotes respectively an arbitrary unary and binary rule.}\label{fig:reductionsteps}
\end{figure}

Traditionally  LK is represented in two-sided sequent calculus. For notational simplicity however, we work in one-sided sequent calculus (Tait-style) with explicit weakening (\weak),  contraction (\con) and permutation (\perm) rules.
Axioms and rules are laid out in~\autoref{tab:calculus} and the cut reduction steps are presented in~\autoref{fig:reductionsteps}.
We generally leave applications of the permutation rule implicit: its only role is to facilitate defining the grammar in the next section.

We use $\alpha$, $\beta$, etc.~for free and $v$, $w$, etc.~for bound variables. 
Upper-case Roman letters, $A$, $B$, etc.\ denote formulæ and upper-case Greek letters $\Gamma$, $\Delta$, etc.\ range over \emph{sequents}, namely finite {sequences} of formulæ. 
We write $\bar A$ to denote the dual of the formula $A$ obtained by de Morgan laws.
A {\em proof} is a finite binary tree labelled by sequents obtained from the axioms and rules of the calculus with the restriction that cuts apply to prenex $\Pi_2/\Sigma_2$ formul\ae{} only.
Without loss of generality, we assume all proofs are {\em regular}, namely strong quantifier inferences are associated unique eigenvariables.
This is particularly relevant in the case of contraction reduction where a sub-proof is duplicated and eigenvariables renamed (expressed by annotating the proof in question by an asterisk) to maintain regularity.
The \emph{length} of a sequent $\Gamma$ is denoted $\size \Gamma$ and 
we  write $\pi\vdash \Gamma$ to express that $\pi$ is a proof with end sequent $\Gamma$.
%%=========================
\section{Proof grammars}\label{sec:grammar_for_Pi2_proofs}
%%=========================
To an LK-proof $\pi\vdash\Gamma$ we associate a typed non-deterministic tree grammar $\Gram G_\pi$ (equivalently, an order-2 recursion scheme) with production rules that abstract the computation of Herbrand sets achieved through Gentzen-style cut-elimination.

Informally, $\Gram G_\pi$ comprises rewrite rules for symbols
$\sigma_{\pi'}^i$ where $\pi' \vdash \Gamma' $ is a sub-proof of $\pi$ and $0 \le i < \size {\Gamma'}$.
The non-terminal $\sigma_{\pi'}^i$ is of function type (of order $\le2$) with arity $\size{\Gamma'}$ returning a sequence of closed terms as witnesses for the weak quantifiers in the $i$-th formula in $\Gamma'$.
The $j$-th argument of $\sigma_{\pi'}^i$ is interpreted as input for the strong quantifiers in the $j$-th formula in $\Gamma'$ and is either a finite sequence (of determined length) or a function from first-order objects to sequences  thereof, the case depending on the quantifier complexity of the corresponding formula.
The type of the arguments to $\sigma_{\pi'}^i$ is independent of~$i$.

As an example, 
consider a derivation $\pi \vdash A_0 , A_1 $ where $A_0$ and $A_1$ are prenex $\Sigma_2 \cup \Pi_2$ formulæ with $m_0$ and $m_1$ existential quantifiers respectively.
The grammar contains two non-terminals associated to $\pi$, $\sigma_\pi^0$ and $\sigma_\pi^1$, of type
\begin{align*}
	\sigma_\pi^0 &: \tau_0 \to \tau_1 \to ( \underbrace{ o \times \dotsm \times o}_{m_0} )
	&
	\sigma_\pi^1 &: \tau_0 \to \tau_1 \to ( \underbrace{ o \times \dotsm \times o}_{m_1} ) 
\end{align*}
where $o$ denotes the type of first-order terms and $\tau_i$ is a type depending on the number of universal quantifiers in $A_i$.
Given terms $T_0 : \tau_0 $ and $T_1 : \tau_1$ the grammar rewrites the term $\sigma_\pi^i T_0 T_1 $ to a sequence of first-order terms $\Seq{t_1,\dotsc,t_{m_i}}$ of length $m_i$ to be interpreted as witnesses to the extensional quantifiers in $A_i$.
The role of $T_0$ and $T_1$ is to provide input for the strong quantifiers (specifically, their corresponding eigenvariables) on which witnesses to the existential quantifiers may depend.
For instance, if $A_i = \forall v \exists w_0 \exists w_1 B_i $ and $B_i$ is quantifier-free for each $i$, then $\tau_0 = \tau_1 = o$, $m_0 = m_1 = 2$ and $\sigma_\pi^i T_0 T_1 $ rewrites to pairs of the form $\Seq{r, s }[\alpha \mapsto T_0][\beta \mapsto T_1]$ where $r$ and $s$ are first-order terms and $\alpha$ and $\beta$ are the eigenvariables for the strong quantifier in $A_0$ and $A_1$ respectively.
Higher-type terms arise in the case of sequents with $\Sigma_2$ formulæ.
Suppose $A_0 = \exists v \forall w B_0$ where $B_0$ is quantifier-free and $A_1$ is as above.
In this case $T_0$ has type $\tau_0 = ( o \to o ) $ and is utilised in generating input for the universal quantifier in $A_0$ modulo witnesses for the existential quantifier.
If the final inference in $\pi$ derives the sequent $ A_0 , A_1 $ from $ \pi_0 \vdash \forall w B_0(v/r), A_1 $ then since
$B_0$ is quantifier-free, the two non-terminals associated to $\pi_0$ are
\begin{align*}
	\sigma_{\pi_0}^0 &:  o  \to o \to o
	&
	\sigma_{\pi_0}^1 &:  o \to o \to ( o \times o ) .
\end{align*}
The production rules corresponding to this inference are
\begin{align*}
	\sigma_\pi^0 T_0 T_1 & \to \Seq{r}
	&
	\sigma_\pi^1 T_0 T_1 & \to \sigma_{\pi_0}^1 (T_0r) T_1.
\end{align*}
The left-hand rule returns the term $r$ as the (single) witness to the existential quantifier in $A_0$ whereas the right-hand rule records the fact that in a witness to the existential quantifier in $A_1$, any occurrence of the eigenvariable for the universal in $A_0$ will be substituted for $T_0r$.
In general, $T_0$ will itself contain non-terminals for other parts of the (wider) proof and will have been introduced through the nesting of non-terminals that occurs when passing through a cut rule (see \autoref{tab:PDrules}).

In the following, fix an LK-proof $\pi \vdash \Gamma$ in which all formulæ are prenex $\Pi_2$/$\Sigma_2$.

\begin{figure}[!b]
  \centering
  \begin{tabular}{|cl|}
	\hline
	Rule of inference & Corresponding production rule(s)
	\\
	\hline
	&\\[-1em]
	$\pi \vdash \bar A,A$
	&
		$\sigma_\pi^i z_0 z_1 \to z_{1-i}$
		\\[0.5em]
	$\vlderivation{
	  \vlin{\forall}{}{\pi\vdash \forall v A,\Gamma}{
          \vlhy{\pi_0\vdash A(v/\alpha),\Gamma}}}$	
	&
	$\sigma_\pi^i ( z_0 \star z_1 )\vec y \to
	(\sigma_{\pi_0}^i  z_1 \vec y)[\alpha\mapsto z_0]$
	\\[1em]
	$\vlderivation{
	  \vlin{\exists}{}{\pi\vdash \exists v A,\Gamma}{
          \vlhy{\pi_0\vdash A(v/r),\Gamma}}}$	
	&
	$\sigma_\pi^i z \vec y \to
	\begin{cases}
	  r \star (\sigma_{\pi_0}^0 (z\cdot r)\vec y),&\text{if $i=0$,}
	  \\
	  \sigma_{\pi_0}^i (z \cdot r)\vec y,&\text{otherwise.}
	\end{cases}$
	\\[1.5em]
	$\vlderivation{
	  \vliin{\cut}{}{\pi\vdash \Gamma,\Delta}{
          \vlhy{\pi_0\vdash A,\Gamma}}{
          \vlhy{\pi_1\vdash \bar A,\Delta}}}$
	&
	$\sigma_\pi^i \vec x \vec y
	\to
	\begin{cases}
	  \sigma_{\pi_0}^{i+1}((\sigma_{\pi_1}^0 \circ_{\bar A} \sigma_{\pi_0}^0)\vec y\vec x)\vec x ,&\text{$ i < \size \Gamma$,}
	  \\
	  \sigma_{\pi_1}^{i'+1} ((\sigma_{\pi_0}^0 \circ_A \sigma_{\pi_1}^0 )\vec x\vec y)\vec y, &\text{$i \ge \size \Gamma$, $i'=i-\size \Gamma$.}
	\end{cases}$
	\\[1.5em]
$\vlderivation{
	  \vlin{\con}{}{\pi\vdash A,\Gamma}{
          \vlhy{\pi_0\vdash A,A,\Gamma}}}$
	&
	$ \sigma_\pi^i z \vec y \to
	\begin{cases}
		\sigma_{\pi_0}^{0} zz\vec  y \mid \sigma_{\pi_0}^{1} zz\vec  y,&\text{if $i=0$,}
	  \\
	  \sigma_{\pi_0}^{i+1} zz\vec  y,&\text{$0<i < \size \Gamma$.}
	\end{cases}$
	\\[0.5em]
	$\vlderivation{
	  \vlin{\weak}{}{\pi\vdash A,\Gamma}{
          \vlhy{\pi_0\vdash \Gamma}}}$
	&
	$
	\sigma_\pi^i z\vec y \to
	\begin{cases}
% 	  \Seq{c_1,\dotsc,c_{u(A)}},&\text{for $i=0$ and fresh constant symbols $(c_j)$,}
		\mathsf c_{A}, &\text{for $i=0$,}
	  \\
	  \sigma_{\pi_0}^{i-1}\vec y,&\text{otherwise.}
	\end{cases}
	$
	\\[1em]
	$\vlderivation{
	  \vlin{{\perm}}{}{\pi\vdash\Gamma,A, B,\Delta}{
          \vlhy{\pi_0\vdash\Gamma,B,A,\Delta}}}$
	&
	$
	\sigma_\pi^i\vec x z_0  z_1 \vec y \to
	\begin{cases}
	  \sigma_{\pi_0}^{i+1}\vec x z_1  z_0 \vec y
	  , &\text{if $ i=\size \Gamma$,}
	  \\
	  \sigma_{\pi_0}^{i-1}\vec x z_1  z_0 \vec y
	  , &\text{if $i = \size \Gamma+1$,}
	  \\
	  \sigma_{\pi_0}^{i}\vec x z_1  z_0 \vec y
	  ,&\text{otherwise.}
	\end{cases}
	$
	\\\hline
  \end{tabular}
  \caption{Production rules: $\vec x$ and $\vec y$ denote sequences of distinct variables of length $\size\Gamma$ and $\size \Delta$ respectively;
	  the contraction rule is the only inference introducing non-determinism; $\mathsf c_k$, $\circ_F$ and $z\cdot r$ are abbreviations for terms described in Section~\ref{sec:production_rules}.}
  \label{tab:PDrules}
\end{figure}
%%=========================
\subsection{Terms and types}\label{sec:types}
%%=========================
We expand first-order terms by a form of explicit substitution, resulting in  \emph{structured (first-order) terms}:
every first-order term is a structured term, and if $s$ and $t$ are structured terms and $\alpha$ is a free-variable symbol then the expression $s [\alpha \mapsto t]$ is a structured term.

Let $o$ denote the type of structured first-order terms and $\epsilon$ the unit type with a single element $\emptyseq : \epsilon$.
$o$ and $\epsilon$ are called \emph{ground-types} and their elements \emph{ground-terms}.
We consider the explicit substitution constructors above as term building operations on both $o$ and $\epsilon$.
A type hierarchy is formed by closing the ground-types under the usual pair-types and function-types: given $u,x:\rho$ and $u':\rho'$ we have $u \star u' : \rho \times \rho'$ and $\lambda x u' : \rho \to \rho'$.
To avoid unnecessary parenthesis the three binary (infix) operations are assumed to associate to the right.
We define $o^0 = \epsilon$ and $o^{k+1} = o \times o^k$.
An element of a type of the form  $o^k $ is called a \emph{sequence-term}.
Given a (possibly empty) sequence $(u_i :o)_{i < k}$ of terms of type $o$, we write $\Seq{u_0, \dotsc, u_{k-1}}$ to abbreviate the sequence-term $u_0 \star \dotsm \star u_{k-1} \star \emptyseq $ of type $o^k$.

Let $\Gamma=\{A_0,\dotsc,A_n\}$.
The type of $\sigma_{\pi}^i$ is given by
\begin{gather*}
	\sigma_{\pi}^i : \tau^*_{A_0} \to \dotsm \to \tau^*_{A_n} \to \tau_{A_i}
\end{gather*}
where $\tau_F$ and $\tau^*_F$ are determined by the complexity of $F$:
\begin{itemize}
	\item for $F = \forall v_1 \dotsm \forall v_m \exists w_1 \dotsm \exists w_n G$ with $G$ quantifier-free,
\begin{align*}
	\tau_{F}& = o^n
	&
	\tau^*_{F}&= o^m;
\end{align*}
	\item for $F = \exists v_1 \dotsm \exists v_m \forall w_1 \dotsm \forall w_n G$ with $n>0$ and $G$ quantifier-free,
\begin{align*}
	\tau_{F} &= o^m
	&
	\tau^*_{F}&=
		\underbrace{o \to \dotsm \to o}_m \to o^{n}.
\end{align*}
\end{itemize}

The \emph{order} of a type $\rho$ (and term of type $\rho$), $\mathit{ord}(\rho)$,  is defined as usual:
$\mathit{ord}(\epsilon)= \mathit{ord}(o)=0$, $\mathit{ord}(\rho \times \rho')= \max\{\mathit{ord}(\rho), \mathit{ord}(\rho')\}$ and $\mathit{ord}(\rho \to \rho') = \max\{ \mathit{ord}(\rho)+1, \mathit{ord}(\rho') \}$.
Thus for a proof $\pi \vdash \Gamma$ where $\Gamma$ is a set of prenex $\Pi_2$ and $\Sigma_2$ formulæ, and for $i<\size \Gamma$, the order of $\sigma_{\pi}^i$ is no greater than $2$.

In the sequel we avoid explicit mention of types when they can be inferred from context.

%%=========================
\subsection{Production rules}\label{sec:production_rules}
%%=========================
Let $\Sigma$ be a finite set of (typed) variable symbols.
A \emph{structured $\lambda$-term over $\Sigma$} is a well-typed term constructed from ground-terms, variables and non-terminals via the term-forming operations described above in which any freely occurring variable is an element of $\Sigma$.

The production rules for non-terminals are determined by the final rule applied to the index proof and are presented in \autoref{tab:PDrules}.
Each production rule has the form $u \to u'$ where the two terms are of the same ground-type and $u'$ is a structured $\lambda$-term over the free variables in $u$.
With the exception of the $\forall$ production rules (which are discussed below) $u$ has the form $\sigma_\pi^i x_0 \dotsm x_n$
which is often condensed to $\sigma^i_\pi\vec x$.
Note that the contraction rule is the only inference rule that introduces non-determinism.

The presentation of the production rules includes the following abbreviations.
The symbol $\mathsf c_A$ appearing in the production rule for weakening (\weak) denotes the sequence-term $\Seq{\mathsf c, \dotsc, \mathsf c}:o^k$ where $\mathsf c:o$ is some fixed constant symbol and $k$ is the number of existential quantifiers in $A$.

The binary operation $\cdot$ appearing in the rule for existential quantifiers ($\exists$) extends term application to cases in which the first argument has type $\epsilon$ in a trivial way:
\begin{gather*}
	z\cdot x =
	\begin{cases}
		z,	&\text{if $z:\epsilon$,}
		\\
		zx,	&\text{otherwise.}
	\end{cases}
\end{gather*}
Its role is to compensate for the case that $A$ is a $\Sigma_1$ formula whereby $\tau^*_{\exists vA}=\tau^*_{A}=\epsilon$;
otherwise  $\tau^*_{\exists vA} = o \to \tau^*_A$ and so $z\cdot r:\tau^*_{A}$ as required.

In the production rules for cut (\cut), the operation $\circ_F$ abbreviates combining two non-terminals depending on the quantifier complexity of $F$.
Let $\pi_0 \vdash \Gamma$ and $\pi_1 \vdash \Delta$ be LK-proofs, and $i<\size \Gamma$, $j<\size \Delta$.
For non-terminals $\sigma_{\pi_0}^0$ and $\sigma_{\pi_1}^0$ and variable sequences $\vec x$ and $\vec y$ of length $\size \Gamma-1$ and $\size \Delta-1$ respectively, we define
\begin{gather*}
	(\sigma_{\pi_0}^0 \circ_F \sigma_{\pi_1}^0)\vec x\vec y
	=
	\begin{cases}
	  \emptyseq,
		&\text{if $F$ is quantifier-free}
		\\
		\lambda z_0 \dotsm \lambda z_m.\,
		\sigma_{\pi_0}^0
			\Seq{ z_0 , \dotsc, z_m }
			\vec x,
		&\text{if $F=\forall v_0 \dotsm \forall v_m G $ and $ G \in \Sigma_1$,}
		\\
		\sigma^0_{\pi_0}
		(
			\lambda z_0 \dotsm \lambda z_m.\,
			\sigma_{\pi_1}^0
				\Seq{ z_0 , \dotsc , z_m }
				\vec y
			)
			\vec x,
		&\text{if $F = \exists v_0\dotsm \exists v_m G $ and $ G \in \Pi_1$.}
	\end{cases}
\end{gather*}
Observe that according to the typing introduced earlier, the production rules listed in \autoref{tab:PDrules} are well-typed (by definition, the axiom case applies only when $A$ is quantifier-free whence all relevant types are identical).

Before we proceed with the definition of language it is important to address the universal introduction rule $\forall$.
In its stated form, $\Gram G_\pi$ is context-sensitive as the reduction depends on the form of at least one argument.
A context-free grammar can be obtained by formulating the production rules using projection functions for pair-types:
\begin{gather}
  \sigma_\pi^n z \vec y \to
  (\sigma_{\pi_0}^n  (p_1 z) \vec y)[\alpha\mapsto p_0 z].
  \label{eqn:Forall_1}
\end{gather}
Doing so, however, will in general expand the language of the grammar:
 if $T$ is a term that non-deterministically rewrites to $r_0\star T_0$ and $r_1\star T_1$ (and the four terms are pairwise distinct) then substituting $T$ for $z$ in \eqref{eqn:Forall_1} yields four combinations of terms, compared with just two available from the $\forall$ production rule.
Nevertheless, this increase will be always finite.
%
%%=========================
\subsection{Language}\label{sec:language}
%%=========================
A \emph{derivation} in the grammar is a sequence of structured $\lambda$-terms containing non-terminals obtained by applying a sequence of grammar production rules and $\beta$-reductions.
For a well-typed term $u$ (possibly containing non-terminals), $L(u)$ denotes the set of terms derivable from $u$ to which no further rules may be applied.
We write $ u \sim u'$ if $L(u)=L(u')$.

Given a sequence-term $T=\Seq{t_1,\dotsc,t_k}:o^k$ of structured first-order terms, let $T^*$ denote the result of evaluating all explicit substitutions occurring in $T$, forming a sequence of first-order terms.
Fix a sequent $\Gamma=\exists \vec v_0 A_0,\dotsc, \exists \vec v_k A_k$ of prenex $\Sigma_1$ formulæ wherein for each $i \le k$, $A_i$ is quantifier-free and $\exists \vec v_i$ abbreviates a block of existential quantifiers of length $a_i$.
Let $\pi \vdash \Gamma$ be an LK-proof with cuts of complexity at most prenex $\Pi_2$/$\Sigma_2$.
The \emph{language} of $\pi$, denoted $L(\pi)$, is the set of pairs $(i,T^*)$ such that $i\le k$ and $T:o^{a_i}$ is a structured sequence-term free of non-terminal symbols derivable from the term $\sigma_\pi^i\emptyseq \dotsm \emptyseq $.
The next lemma demonstrates that the choice of starting symbol is canonical.

\begin{lemma}\label{lem:1}
	If $\pi\vdash A, \Gamma$ and $A$ is prenex $\Sigma_1$ then for all terms $u _0, \dotsc, u_{\size{\Gamma}}$ of the appropriate type we have $\sigma_\pi^i u_0 \dotsm u_{\size{\Gamma}} \sim \sigma_\pi^i \emptyseq u_1 \dotsm u_{\size{\Gamma}}$.
\end{lemma}

Since the production rules are naturally acyclic (rewriting a non-terminal $\sigma_\pi^i$ introduces only non-terminals indexed by strict sub-proofs of $\pi$), we deduce
\begin{lemma}\label{lem:finite}
  For any regular proof $\pi$, $L(\pi)$ is finite.
\end{lemma}
As a consequence of \autoref{lem:finite} the language of a proof $\pi$ can be viewed as inducing an \emph{expansion} of its end-sequent, obtained by replacing each formula $\exists \vec v_i A_i$ by the corresponding disjunction $\bigvee \{ A_i(\vec t) \mid (i,\vec t)\in L(\pi)\}$.
\begin{theorem}\label{lem:expansion}
  If $\pi \vdash \Gamma$ is an LK-proof of a prenex $\Sigma_1$ sequent in which all cuts are prenex $\Pi_2$ and $\Sigma_2$ formulæ then the expansion of $\Gamma$ induced by $L(\pi)$ is a tautology.
\end{theorem}
The proof of this theorem (and the more general statement in \autoref{thm:Main_Thm}) is covered in Section~\ref{sec:reduction_steps} below by establishing that the language of a proof is preserved through most cut reduction steps.
In the (base) case that all cuts in $\pi$ are on quantifier-free formulæ, we observe that the grammar rules merely associate to each weak quantifier in the end-sequent the witnesses as they appear in $\pi$.
%%=========================
\section{Language preservation}\label{sec:reduction_steps}
%%=========================
Let $\pi\rightsquigarrow \pi'$ express that $\pi'$ is obtained from $\pi$ by the application of a reduction rule in~\autoref{fig:reductionsteps} to a sub-proof of $\pi$.
In the present section we determine for which reduction steps $\pi\rightsquigarrow \pi'$ we have:
(i) \emph{language inclusion}: $L(\pi)\supseteq L(\pi')$; and
(ii) \emph{language equality}: $L(\pi)=L(\pi')$.
Language inclusion will suffice to derive the main theorem;
equality allows a  more fine-grained study of the Herbrand content of proofs as if $\pi_0$ and $\pi_1$ are proofs that can be connected by a sequence of forward and backward language-preserving reduction steps then $L(\pi_0)=L(\pi_1)$.

The structure of our proof grammars is such that to deduce inclusion or equality it suffices to analyse the reduction steps locally:
%%=========================
\subsection{Cut permutation}\label{subs:cut_permutation}
%%=========================
We begin by considering the instances of the binary inference reduction that permute two cuts. Suppose $\pi \rightsquigarrow \pi'$ are the two proofs
\begin{align}\label{eqn:cut_perm_proof}
  \vlderivation
  {
	\vliin{\cut}{}{\pi \vdash \Gamma,\Delta,\Lambda}
	{
	  \vliin{\cut}{}{G , \Gamma, \Delta}
	  {
		\vlhtr {\pi_0} {F , G , \Gamma}
	  }{
		\vlhtr {\pi_1} {\bar F , \Delta}
	  }
	}{
	  \vlhtr   {\pi_2} {\bar G , \Lambda}
	}
  }
  &&\vlderivation
  {
	\vliin{\cut}{}{\pi' \vdash \Gamma,\Delta,\Lambda}
	{
	  \vliin{\cut}{}{F , \Gamma, \Lambda}
	  {
		\vlhtr {\pi_0} {F , G , \Gamma}
	  }{
		\vlhtr {\pi_2} {\bar G , \Lambda}
	  }
	}{
	  \vlhtr   {\pi_1} {\bar F , \Delta}
	}
  }
\end{align}
\begin{lemma}\label{lem:cut_perm_proof1}
  For $\pi$ and $\pi'$ in \eqref{eqn:cut_perm_proof}, if either $F$ or $G$ is $\Sigma_2$ then $L(\pi)=L(\pi')$.
\end{lemma}
\autoref{lem:cut_perm_proof1} covers all cases of permuting two cuts that suffice for establishing \autoref{thm:Main_Thm}.
In the case both $F$ and $G$ are (genuine) $\Pi_2$ formul\ae, the language of the grammars need not be preserved:
\begin{lemma}\label{lem:cut_perm_proof2}
  There are instantiations of $\pi$ and $\pi'$ in \eqref{eqn:cut_perm_proof} such that $L(\pi)$ and $L(\pi')$ are incomparable.
\end{lemma}
%
%%=========================
\subsection{Contraction reduction}\label{subs:contraction_reduction}
%%=========================
Consider the proofs
\begin{align}\label{eqn:cont_red_proof1}
  \vlderivation
  {
	\vliin{\cut}{}{\pi \vdash \Gamma,\Delta}
	{
	  \vlin{\con}{}{F , \Gamma}
	  {
		\vlhtr{\pi_0}{F , F , \Gamma}
	  }
	}{
	  \vlhtr{\pi_1}{\bar F , \Delta}
	}
  }
  &&
  \vlderivation
  {
	\vlin{\con^*}{}{\pi' \vdash \Gamma,\Delta}
	{
	  \vliin{\cut}{}{\Gamma,\Delta,\Delta}
	  {
		\vliin{\cut}{}{F , \Gamma,\Delta}
		{
		  \vlhtr{\pi_0}{F , F , \Gamma}
		}{
		  \vlhtr{\pi_1}{\bar F , \Delta}
		}
	  }{
		\vlhtr{\pi_1^*}{\bar F , \Delta}
	  }
	}
  }
\end{align}
\begin{lemma}
  For $\pi$ and $\pi'$ in \eqref{eqn:cont_red_proof1}, if $F$ is $\Sigma_2$ then $L(\pi)=L(\pi')$.
\end{lemma}
As in the previous case, language inclusion does not hold in general when reducing a contraction.
Specifically, if (i) $F$ is a genuine $\Pi_2$ formula i.e.~$F = \forall v_0\dotsm \forall v_k \exists w G $ for some $\Sigma_1$ formula $G$, and (ii) there are contractions on $\bar F$ in the subproof $\pi_1$ then the languages  $L(\pi)$ and $L(\pi')$ can be incomparable.
We do, however, have
\begin{lemma}
  For $\pi$ and $\pi'$ in \eqref{eqn:cont_red_proof1}, if $F$ is $\Pi_2$ and there are no contractions on $\bar F$ in the sub-proof $\pi_1$ then $L(\pi') \subseteq L(\pi)$.
\end{lemma}
%
%%=========================
\subsection{Quantifier reduction}\label{subs:quantifier_reduction}
%%=========================
\begin{lemma}
   For $\pi$ and $\pi'$ in \eqref{eqn:qunt_red_proof1} we have $L(\pi)=L(\pi')$.
\end{lemma}
\begin{align}\label{eqn:qunt_red_proof1}
  \vlderivation
  {
	\vliin{\cut}{}{\pi \vdash \Gamma,\Delta}
	{
	  \vlin{\forall}{}{\forall v\, F , \Gamma}
	  {
		\vlhtr{\pi_0}{F(v/\alpha) , \Gamma}
	  }
	}{
	  \vlin{\exists}{}{\exists v\, \bar{F} , \Delta }
	  {
		\vlhtr{\pi_1}{\bar{F}(v/t) , \Delta }
	  }
	}
  }
  &&
  \vlderivation
  {
	\vliin{\cut}{}{\pi' \vdash \Gamma,\Delta}
	{
	  \vlbtrf{\pi_0^ { [\alpha\mapsto t] } }{F(v/t) , \Gamma}{6em}{0.6}
	}{
	  \vlbtrf{\pi_1}{\bar{F}(v/t) , \Delta }{2em}{1}
	}
  }
\end{align}
%%=========================
\subsection{Quantifier permutation}\label{subs:quantifier_permutation}
%%=========================
Consider permuting a universal quantifier with a cut:
\begin{align}\label{eqn:quant_perm_proof1}\vlderivation
  {
	\vliin{\cut}{}{\pi \vdash \forall vA , \Gamma,\Delta}
	{
	  \vlin{\forall}{}{\forall v A , \Gamma,F}
	  {
		\vlhtr{\pi_0}{A(v/\alpha) , \Gamma,F}
	  }
	}{
	  \vlhtr{\pi_1}{\Delta , \bar F}
	}
  }
  &&
  \vlderivation
  {
	\vlin{\forall}{}{\pi' \vdash \forall v A , \Gamma,\Delta}
	{
	  \vliin{\cut}{}{A(v/\alpha) , \Gamma,\Delta}
	  {
		\vlhtr{\pi_0}{A(v/\alpha) , \Gamma , F}
	  }{
		\vlhtr{\pi_1}{\Delta,\bar{F}}
	  }
	}
  }
\end{align}
\begin{lemma}\label{lem:Quant_Perm}
   For $\pi$ and $\pi'$ in \eqref{eqn:quant_perm_proof1} we have  $L(\pi')\subseteq L(\pi)$.
\end{lemma}
%%=========================
\subsection{Remaining reductions}\label{subs:remaining_reductions}
%%=========================
The remaining rules are straightforward to analyse and all induce language equality except for weakening reduction for which we have language inclusion.
%%=========================
\section{Conclusion}\label{sec:conclusion}
%%=========================
To each proof in first order logic with prenex $\Pi_2/\Sigma_2$ cuts we associate a formal grammar abstracting the semantic aspect of cut elimination and classify the cut reduction and permutation rules according to whether or not the language of the grammar is preserved under these rules.
The ultimate goal of the study is to extend this classification to arbitrary classes of cut-formulæ.

The grammars utilised in this paper have a number of advantages over previous language-theoretic approaches for proofs with $\Pi_2/\Sigma_2$ cuts \cite{AHL15,AHL16}.
We can now deal with non-simple proofs (i.e.\ proofs admitting contractions on universal formul\ae) as well as blocks of like quantifiers.
Furthermore, unlike the grammars devised in~\cite{AHL15,AHL16}, derivations are not restricted by equality constraints (or rigidity requirements).
As a result, preservation of language over the cut reduction steps reduces more or less to mere computation.
Also notable is the fact that the production rules of the grammar given here are both acyclic and unidirectional relative to the associated proof:
each production rule rewrites a non-terminal in favour of non-terminals labelled by strict sub-proofs.
Finally, the new grammars make the appearance of non-confluence in cut-elimination more transparent.
\paragraph*{Acknowledgements}The authors’ research was supported by the Wiener Wissenschafts-, Forschungs- und Technologiefonds (WWTF),  project no.~VRG12-04. The authors wish to thank the anonymous referees for their helpful comments and suggestions.
%%=========================%
\bibliography{references}
%%=========================
\end{document}